# Anisotropic exciton polariton pairs as a platform for PT-symmetric non-Hermitian physics

Devarshi Chakrabarty[1†], Avijit Dhara[1†], Pritam Das[1], Kritika Ghosh[2], Ayan Roy Chaudhuri[2], Sajal Dhara[1*]

[1]Department of Physics, IIT Kharagpur, Kharagpur-721302, India

[2]Materials Science Centre, IIT Kharagpur, Kharagpur-721302, India

[†] *These authors contributed equally*
[*] *Corresponding author. email: sajaldhara@phy.iitkgp.ac.in*

**Abstract:** Non-Hermitian systems with parity-time (PT) symmetry have been realized using optical constructs in the classical domain, leading to a plethora of non-intuitive phenomena[1–4]. However, PT-symmetry in purely quantum non-Hermitian systems like microcavity exciton-polaritons[5–9] has not been realized so far. Here we show how a pair of nearly orthogonal sets of anisotropic exciton-polaritons can offer a versatile platform for realizing multiple spectral degeneracies called Exceptional Points (EPs) and propose a roadmap to achieve a PT-symmetric system. Polarization-tunable coupling strength[10,11] creates one class of EPs, while Voigt EPs[12,13] are observed for specific orientations where splitting of polariton modes due to birefringence is compensated by Transverse Electric (TE) -Transverse Magnetic (TM) mode splitting. Thus, paired sets of polarized anisotropic microcavity exciton-polaritons can offer a promising platform not only for fundamental research in non-Hermitian quantum physics and topological polaritons, but also, we propose that it will be critical for realizing zero threshold lasers.

**Introduction**

The parameter space of non-Hermitian systems contains spectral singularities called Exceptional Points (EPs) where eigenvalues and their corresponding eigenvectors coalesce[14,15]. Their exploration offers a window into several intriguing phenomena unique to non-conservative systems[16–22]. Furthermore, the balancing of gain and loss allows a non-Hermitian system to become PT-symmetric, in which case the eigenvalues can be real[23]. Near the EPs, tuning the parameters carefully causes a transition from PT-symmetric to PT-broken phase with complex eigenvalues. Realizing this using photonic systems in the classical regime has led to interesting phenomena like loss-induced or single-mode lasing[24–30], non-reciprocity or switching of chiral modes in waveguides[31–35] and Brillouin lasers with bistable memory[36]. On the other hand, a PT-symmetric state in a purely quantum system like cavity exciton-polaritons is yet to be realized due to the coupled exciton and photonic modes both being inherently lossy.

In this work, we strongly couple linearly polarized excitons in an optically biaxial semiconductor, multilayer Rhenium disulphide ($ReS_2$), with photons using a microcavity, creating a pair of polarized anisotropic exciton-polariton sets. The microcavity structure itself is rendered optically biaxial, lifting the degeneracy between the transverse-electric (TE) and transverse-magnetic (TM) cavity modes. Thus, we observe the signature of four polariton branches in angle-resolved reflectance measurements, resulting from the coupling between the two primary exciton species in $ReS_2$[37,38] and the anisotropic cavity's TE and TM modes. We find that by rotating the polarization of incident white light, the interaction strength between the two exciton and cavity modes can be changed from the strong to weak light-matter coupling regime. We model this with a 4x4 Hamiltonian, and find our system shows a peak coupling strength of ~18 meV. The ability to tune coupling strength, as well as energy detuning with incident angle due to the cavity modes' finite dispersion, allows for the realization of four sets of EP pairs in this system. In addition, rotating the orientation of the crystal with respect to the laboratory frame also tunes the polariton dispersion due to the change in the refractive index corresponding to the TE and TM cavity modes. For specific orientations of the sample, we discover another class of EPs known as Voigt points where the polariton branches corresponding to different polarizations coalesce. Thus, we find a robust way to realize exceptional points in a purely quantum system, and even tune their location in the dispersion of the polarized polariton modes. Crucially, the optically active material ReS$_2$ has a large number of Raman modes due to its low symmetry[39], which when pumped resonantly at an EP, can introduce the gain mechanism required to realize a PT-symmetric system. This offers the potential system to realize a zero threshold polariton Raman laser, which we experimentally realize in another work[40]. This shows the ReS$_2$ microcavity system can be a viable platform for observing EP-related phenomena like loss-controlled zero-threshold lasing, non-trivial topology, and PT-symmetry breaking.

## Results

### Anisotropic optical microcavity

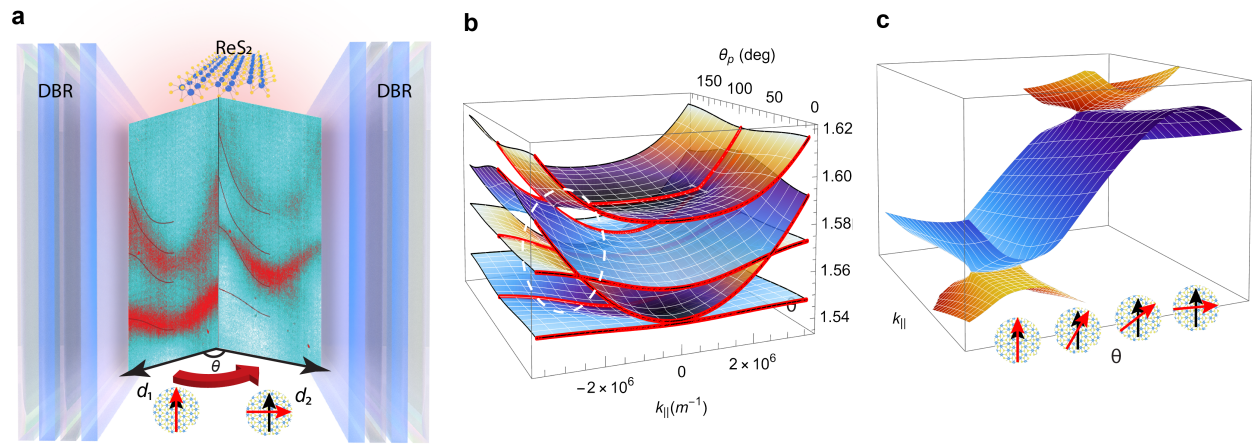

**Fig 1. Exceptional point (EP) pair realized through polarization-dependent coupling.**
**a.** Angle resolved reflectance showing a pair of exciton-polariton dispersions for probe beam polarized along $\widehat{d_1}$ (left half) and $\widehat{d_2}$ (right half), indicated by red arrows (black arrow indicates the b-axis). **b.** Polariton energy dispersions in two-dimensional parameter space of in-plane momentum and polarization angle $\theta$. The red curves mark the polariton branches for $\theta = 0°$ and $90°$ **c.** Zoomed-in view of the dispersion region marked with dotted white ellipse in **(b)** clearly showing the pair of EPs that are probed by changing $\theta$.

Fig 1(a) shows a schematic of the microcavity used in this work, where the optically active 10 nm ReS$_2$ multilayer is embedded between two distributed Bragg reflectors (DBRs). The bottom mirror consists of 10 pairs of SiO$_2$/Ta$_2$O$_5$, while the top mirror has 8 pairs to allow for preferential out-coupling of light from the cavity in the direction of collection. The center wavelength $\lambda_{DBR}$ of the DBRs has a gradient across the substrate due to varying deposition rate of the dielectrics. The ReS$_2$ sample in this case is embedded where $\lambda_{DBR}$ is approximately 735 nm. ReS$_2$ is known to harbor two exciton species, X$_1$ and X$_2$, with specific angles with respect to the crystallographic b-axis. Due to the optical biaxiality of ReS$_2$, light propagating along the z-direction will encounter two different refractive indices based on the electric field polarization direction. This leads to two separate cavity modes, based on different refractive indices $n_c$ corresponding to the two different principal optical axes for propagation along the z-direction. To a very good approximation, their dispersion can be written as:

$$\hbar\omega(k) = E_{cav} + \frac{\hbar^2 k^2}{2m_{ph}}; \quad E_{cav} = \frac{\hbar c \pi}{n_c L_c}, \quad m_{ph} = \frac{\hbar \pi n_c}{c L_c}$$

The cavity length $L_c$ in this expression as well as the refractive index $n_c$ needs to be modified for the cavity with an anisotropic optical medium embedded inside. $L_c$ in this case is $\frac{\lambda_{cav}}{2n_{SiO2}} + L_{ReS2}$, while $n_c$ can be taken to be the length-weighted average, $\frac{n_{SiO2}L_{SiO2} + n_{ReS2}L_{ReS2}}{L_{SiO2} + L_{ReS2}}$, with $n_{ReS2}$ being different for TE and TM polarizations. In addition, since $E_{cav}$ is detuned from the central energy of the DBRs forming the microcavity due to the presence of ReS$_2$, TE-TM splitting must also be considered. In this case, where $E_{cav} > E_{DBR}$, the TM mode will show larger dispersion than TE at higher incident angles[41,42].

**Polarization angle tuned cavity polariton dispersion**

To perform incident polarization-resolved reflectance, white light from a broadband halogen source is first linearly polarized and sent through a half-wave plate (HWP), which can be rotated to change the angle of incident linear polarization $\theta$. As shown in Supplementary Figure S1, rotating $\theta$ changes the relative amplitudes of the TE and TM electric field incident on the sample. The experimental setup is arranged such that the spectrometer detects the TM mode reflectance when $\theta = 0°$. The in-plane components of the TE and TM mode are thus $E\sin\theta$ and $E\cos\theta\cos\phi$, where $\phi$ is the angle of incidence. Since a 60x objective lens with 0.7 NA is used to focus the beam onto the sample, $\phi$ ranges from 0 to 45°. The back focal plane of the objective lens is imaged to obtain a one-shot in-plane momentum vs energy dispersion of the reflected light. Vignetting due to the limited aperture of the tube lens collecting the reflected light, causes the effective observation of dispersion up to reflectance angle $\phi \sim 24°$.

The results of the measurements are plotted in Supplementary Figure S2, where we find clear evidence of strong light-matter coupling, with the two exciton resonances $X_1$ and $X_2$ coupling with the two anisotropic cavity modes to form four polariton modes. All measurements were taken at 4K, with the sample mounted in a closed-cycle Helium cryostat. As $\theta$ is changed, up to four resolvable polariton branches can be seen in the reflectance. For each of the values of $\theta = 60°$ and $140°$, the absorption from a different pair of branches is maximized. This is due to the linearly polarized excitons' oscillator strengths being maximum in those directions, with 60° corresponding to X1 and 140° to X2. These angles, denoted as $\theta_1$ and $\theta_2$, are determined by the sample orientation in the lab frame.

The observed polariton dispersions can be modelled using a four-body coupled oscillator Hamiltonian, expressed here in the basis of creation and annihilation operators, $C_i^\dagger$ denoting the photon creation operator and $X_j^\dagger$ the exciton creation operator:

$$H = C_i C_i^\dagger \hbar\widetilde{\omega}_i + X_j X_j^\dagger \hbar\widetilde{\omega}_j + \left[C_i X_j^\dagger + C_i^\dagger X_j\right] v_j \widehat{e}_i \cdot \widehat{d}_j \quad \ldots\ldots (1)$$

Summation over the indices $i$ and $j$ is implied. $\widehat{e}_i$ denotes the unit vector along the two electric field modes, with $i$=s (p) for TE mode (TM mode), and $\widehat{d}_j$ indicates the unit vector along which the absorption from each of the linearly polarized exciton species is maximum, with $j$=1,2 for $X_1$ and $X_2$ exciton species respectively. $\widetilde{\omega}_i$ and $\widetilde{\omega}_j$ denote the complex angular frequencies of the bare cavity mode and excitons respectively, with their imaginary part being the half-width half maximum linewidths. The inclusion of these imaginary parts, or losses, in the Hamiltonian renders it non-Hermitian.

Since the two exciton species are at ~100° to each other rather than completely orthogonal, the TE and TM modes of the electric field in the cavity couple to both the species of excitons. This is reflected in the off-diagonal terms of the Hamiltonian matrix, $g_{ij} = v_j \widehat{e}_i \cdot \widehat{d}_j$, which represent the various coupling strengths. All the four types of coupling in this system have an explicit dependence on $\theta$ (e.g., for the coupling of X1 with TM mode, $g_{p1} = v_1 \widehat{e_p} \cdot \widehat{d_1} = v_1 cos\theta cos\phi cos\theta_1$ and so on).

Simultaneously fitting the model with the observed polariton dispersions at $\theta = \theta_1$ and $\theta_2$ yields the coupling strength amplitudes $v_1$ and $v_2$ to be 14.5 and 18.6 meV respectively. The dispersion fits are shown as white curves superimposed on the experimental reflectance in Fig S2. The coupling strengths $g_{ij}$ are tuned from their maximum values of (e.g., $v_1 cos\theta_1$ for $g_{p1}$) to zero by changing $\theta$, which results in a transition from strong to weak light-matter coupling between the four oscillator modes in this system. Further, the energy detuning between the uncoupled modes changes with in-plane momentum $k_∥$ or incidence angle $\phi$ because of the inherent dispersion of the cavity modes. Being able to explore this parameter space offers the possibility of realizing EPs in this system (see Supplementary Note 1).

Fig 1(b) shows a 3D plot of the real eigenenergy surfaces for the four-body coupled oscillator system, in the parameter space of $\theta$ and $k_∥$, using parameter values extracted from fitting as mentioned previously. As expected from the model, near $\theta = 90°$ the TM mode electric field component and therefore the corresponding coupling strength $g_{p2} \propto \cos\theta$ approaches zero, falling below the threshold for strong coupling. At $k_∥ = \pm 2.2$ μm$^{-1}$ where the TM mode and X2 exciton have zero energy detuning, the anti-crossing is removed, and a pair of EPs are realized as $\theta$ is tuned to around 90°. These are marked as blue dots on the energy surfaces in Fig 1(c). The corresponding position of the EPs in the experimental reflectance data is shown in Fig S3(a), by taking the cross-section at $k_∥ = \pm 2.2$ μm$^{-1}$ for the reflectance data taken for $\theta$ varying from 0° to 360°.

Similarly, the TE mode-related component $g_{s2} \propto \sin\theta$ approaches zero at $\theta = 0°$ and 180°. At $k_∥ = \pm 2.6$ μm$^{-1}$, the TM mode and X2 energy detuning is such that another pair of EPs is created. These are marked as red dots in Fig 1(c) and indicated in the experimental cross-section data in Fig S3(b) with a dotted circle. The reduced absorption by the polariton branches near the EPs make the coalescing difficult to observe in the experimental reflectance, but their presence can be surmised from the change in coupling strengths in the Hamiltonian, and extrapolation of the experimental data.

**Revealing Voigt Exceptional Points in $k_x$-$k_y$ parameter space**

The orientation of the sample b-axis in the laboratory frame is crucial in deciding the occurrence and nature of EPs in the system. Essentially $\theta_1$ and $\theta_2$, the orientations of the excitons with respect to the TM cavity mode which is aligned along the laboratory x-axis, are decided by the orientation of the mounted sample. In addition, the dispersion of the TE and TM modes are changed since the refractive index corresponding to these polarization directions is also changed with orientation.

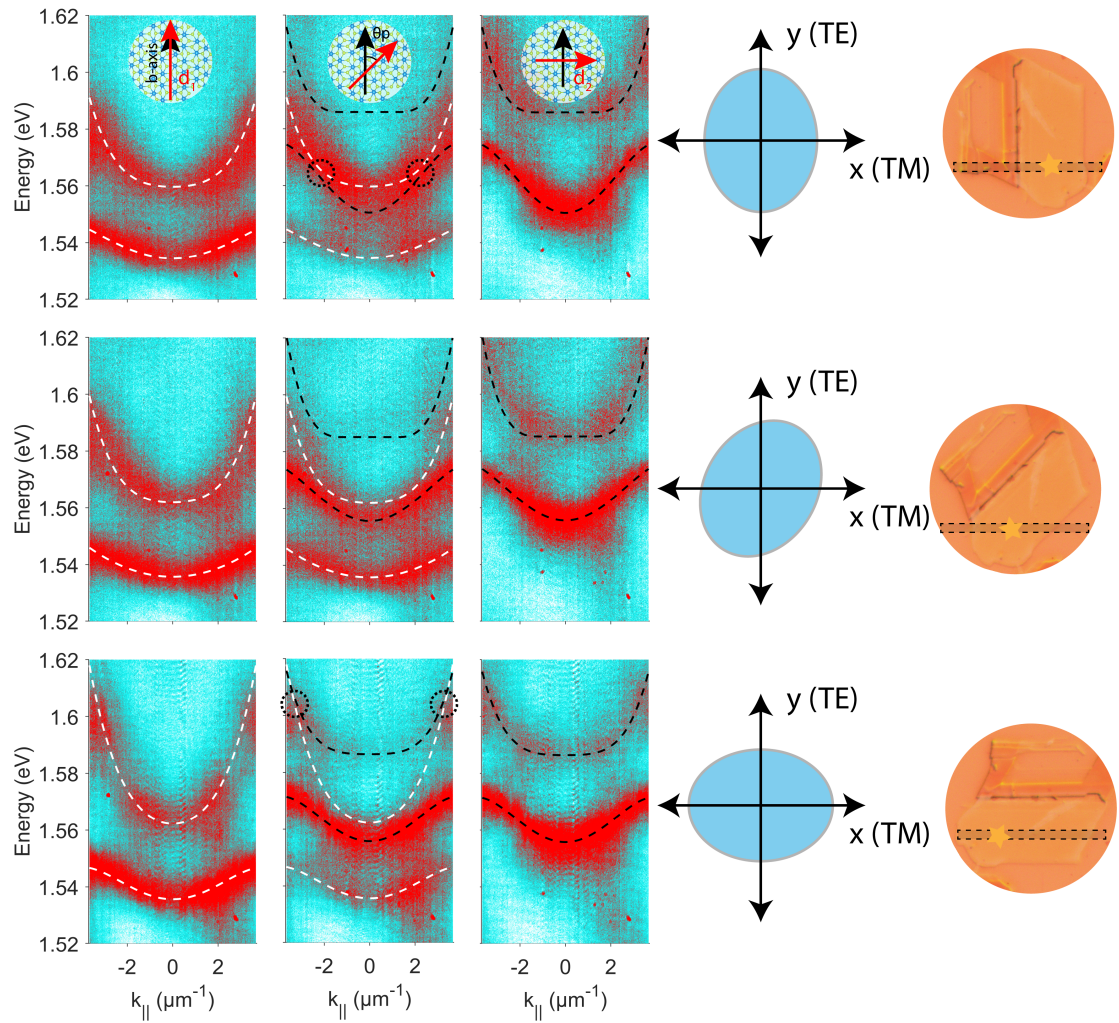

**Fig 2. Voigt EPs revealed in sample orientation-dependent polariton dispersions.** First three columns show the angle resolved reflectance for probe beam polarization along $\widehat{d_1}$, intermediate, and $\widehat{d_2}$ as indicated by the insets in the topmost row. The intermediate case reveals the Voigt EPs where the polarized pair of polariton modes coalesce (marked by black dotted circles). Last two columns represent the corresponding refractive index ellipsoid and the sample orientation with respect to the laboratory frame. The projection of the spectrometer slit is marked as a black dotted rectangle on the sample image.

In the previous section examining polarization-tuned coupling, $\theta_1$ was fixed at 60°. In Fig 2, the angle-resolved reflectance data for sample orientations $\theta_1$= 0°, 45° and 90° are shown for when the incident polarization is kept along $\widehat{d_1}$, an intermediate direction, and $\widehat{d_2}$. For polarization along $\widehat{d_1}$, only two polariton branches are 'probed' or visible as absorption lines. We denote the upper and lower polariton associated with the $\widehat{d_1}$ direction as $U_1$ and $L_1$ respectively. Similarly, for incident polarization along $\widehat{d_2}$ we observe two polariton branches corresponding to $X_2$, thus denoted $U_2$ and $L_2$. In the intermediate incident polarization state, all four branches are observed, and we find for specific orientations the polariton branches coalesce at specific values of in-plane momentum, or incidence angle. These points represent a class of EPs known as Voigt points. In this system, this is achieved when the difference in energy between TE and TM modes at zero incidence angle caused by birefringence is compensated by the TE-TM splitting at higher angles. For $\theta_1$= 0°, we find a pair of Voigt EPs at ~1.56 eV and $k_{||}$=1.8 µm$^{-1}$ where the branches $U_1$ and $L_2$ coalesce. For $\theta_1$= 90°, Voigt EPs are formed when $U_1$ and $U_2$ coalesce, at ~1.6 eV and $k_{||}$=3.5 µm$^{-1}$.

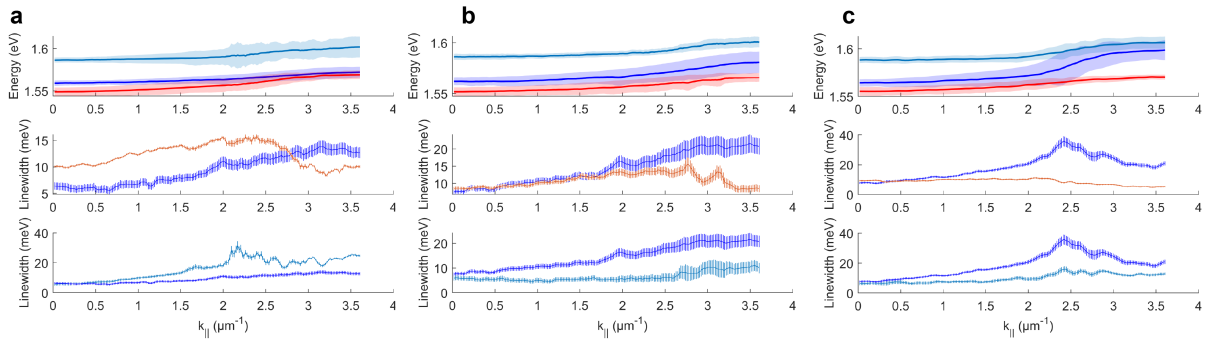

**Fig 3. Extracted energies and linewidths (FWHM) of polariton branches for different sample orientations.** The topmost plot shows the extracted dispersions of $L_2$ (red), $U_1$ (blue) and $U_2$ (light blue) polaritons for sample orientation - **a.** $\theta_1 = 0°$, **b.** $\theta_1 = 45°$, **c.** $\theta_1$= 90°, with the color spread indicating the linewidth. The corresponding polariton linewidths are plotted with the same color in the middle and bottom panels. The error bars represent 95% confidence intervals of the fitted value. For $\theta_1 = 0°$, we see a clear crossing of the linewidths of $U_1$ and $L_2$ near the EP where the real eigenvalues also coalesce.

To probe the complex eigenvalues of the polaritons, we carefully extract the position and linewidths of the polariton branches for incident polarization along $\widehat{d_1}$ and $\widehat{d_2}$ directions, by fitting the absorption dips with Lorentzian functions. The results are shown in Fig 3. In Fig 3(a) corresponding to sample orientation $\theta_1 = 0°$, we find that the real values of $U_1$ and $L_2$ branches coalesce, and in the same $k_\parallel$ range the linewidths show a crossing, clearly revealing an EP where the complex eigenvalue of both modes have coalesced. In Fig 3(b) where $\theta_1 = 45°$, no EPs are observed, with the polariton branches showing level repulsion characteristic of strong coupling while the linewidths also don't show any crossing. For $\theta_1 = 45°$ in Fig 3(c) we find that while the real values of $U_1$ and $U_2$ coalesce at a high $k_\parallel$, we see an anti-crossing in the linewidths. This indicates that the critical condition for realizing an EP is not achieved in this case, and the coupling strength is such that the system is in a PT-broken phase. The bottom panel shows the linewidth of $U_1$ and $U_2$ are approaching each other, so a Voigt point may form at a higher $k_\parallel$ than which can be probed with our current setup. This possibility is explored by using our 4x4 Hamiltonian model to simulate the polariton dispersions.

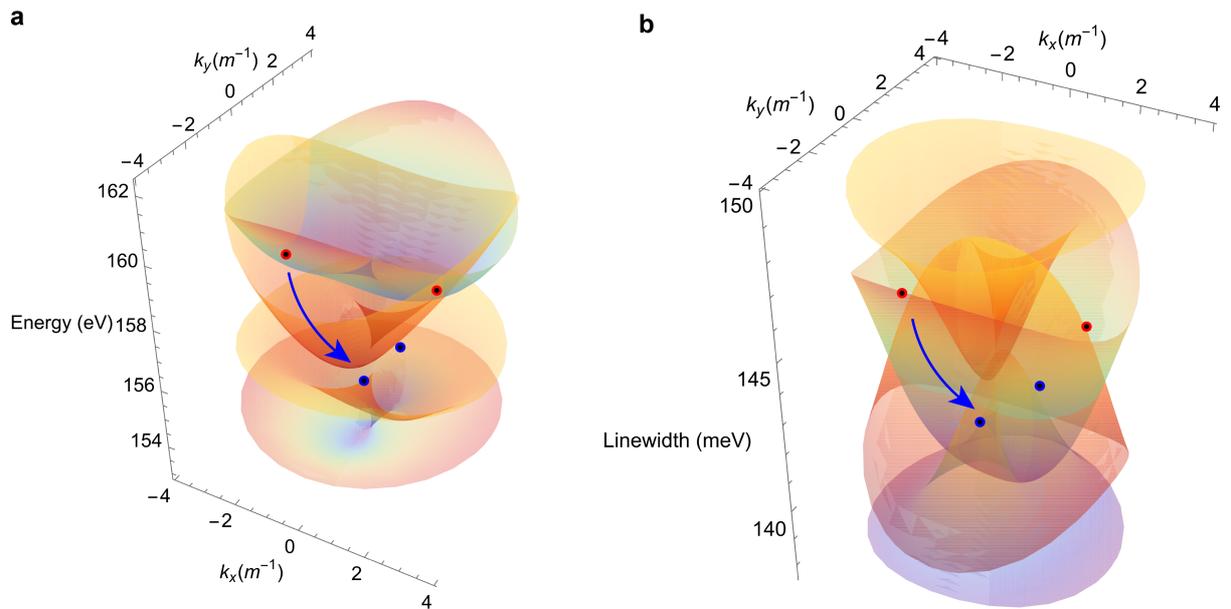

**Fig 4. Polariton pair dispersion in $k_x$- $k_y$ parameter space. a, b** Real and imaginary parts of the eigenenergy of the 4x4 Hamiltonian, as a function of $k_x$ and $k_y$. Two sets of Voight EPs are marked

with red and blue dots where the eigenvalues coalesce. Blue arrow indicates a trajectory to adiabatically tune the system from one EP to another by varying $k_y$.

A co-ordinate transformation of the 4x4 Hamiltonian given in Eqn (1) by fixing the value of $\theta$ and considering only the change in $\theta_1$ allows us to go from $k_{||} - \theta$ space to $k_x$-$k_y$ space which reveals the sample orientation or $\theta_1$ dependent dispersion of the polaritons, shown in Fig 4. In this case, the refractive index corresponding to the TE or TM mode is $\theta_1$ dependent, which we model as $n_{c1} = \sqrt{(n_{cx}cos\theta_1)^2 + (n_{cy}sin\theta_1)^2}$ and $n_{c2} = \sqrt{(n_{cx}sin\theta_1)^2 + (n_{cy}cos\theta_1)^2}$ which represents the index ellipsoid of the optically active biaxial ReS$_2$ for propagation in the z-direction. $n_{cx}$ and $n_{cy}$ are extracted first by fitting the polariton branches for the cases when $\theta_1 = 0°\ and\ 90°$ respectively. In Fig 4(a) we find two sets of Voigt EPs revealed, as found in our experimental results, one set between U$_1$ and L$_2$, and one set between U$_1$ and U$_2$. Fig 4(b) shows the corresponding linewidths that also show crossing near the Voigt points. On changing the parameter $\theta_1$ or $k_{||}$, the state of the system can be quasi-adiabatically tuned and encircled around an EP, which will cause a 'flipping' of the state where the real value stays the same while the imaginary part changes after one encirclement.

**Prescription for realizing a PT-symmetric system**

The uncoupled cavity and exciton modes in Eqn (1) have an inherent lifetime after which they decay, signified by a negative sign of their complex energy values. This property is inherited by the polariton modes which are also lossy and have a finite lifetime. To make the system PT-symmetric, a mechanism to introduce gain in to one of the modes is required. If a low-symmetry material with many Raman modes is embedded in the cavity, it becomes possible to pump the cavity resonantly at one polariton branch and ensure one of the Raman Stokes lines coincides with another polariton branch at a lower energy. By changing the pump power, it is possible to tune the gain or loss of the resonant polariton-Raman mode state and thus achieve PT-symmetry. This manifests as lasing associated with the Raman mode resonant with a polariton branch. Further, the EPs in this system as well as the polarized cavity modes can be exploited to achieve thresholdless lasing. By pumping at an EP, both polarization states are excited, while only the Raman polarization state that is resonant with a cavity mode is enhanced and starts lasing. Since only one mode participates in the lasing process (instead of all polarizations in an isotropic microcavity), the result is truly thresholdless lasing. The experimental realization of this is discussed in detail in our other work[40].

**Discussion**

In summary, the pair of interacting linearly polarized exciton-polaritons realized in this work offer an excellent platform to observe non-Hermitian phenomena in a purely quantum system due to the direct control over coupling strength. Probe polarization-resolved reflectance and sample rotation offers a window into the presence of two classes of EPs in the polariton dispersion. Due to the quasi-indirect bandgap of $ReS_2$, the photoluminescence (PL) quantum yield is poor and the signal near the EPs is extremely weak. Future work can attempt to investigate non-Hermitian topological phases which can be found in the PL from the polaritons near the EPs. More importantly, we find that birefringent crystals like $ReS_2$ which have several Raman modes open up a way to realize PT-symmetry in a non-Hermitian polariton system, which is essential for realizing zero threshold lasers.

**Acknowledgements:**

This work has been supported by funding from the Science and Engineering Research Board (CRG/2018/002845, CRG/2021/000811); Ministry of Education (MoE/STARS- 1/647); Council of Scientific and Industrial Research, India (09/081(1352)/2019-EMR-I); Department of Science and Technology, Ministry of Science and Technology, India (IF180046) and Indian Institute of Technology Kharagpur.

**Author contributions:**

SD conceptualized the project. DC, AD, KG, ARC and SD contributed in the sample fabrication. DC, AD, PD and SD formulated the experiments, performed data analysis, and developed the theory. DC, AD, performed all optical measurements. DC, AD, and SD drafted the paper, and all authors contributed to reviewing and editing the final draft. SD supervised the project.

**Supplementary Materials**

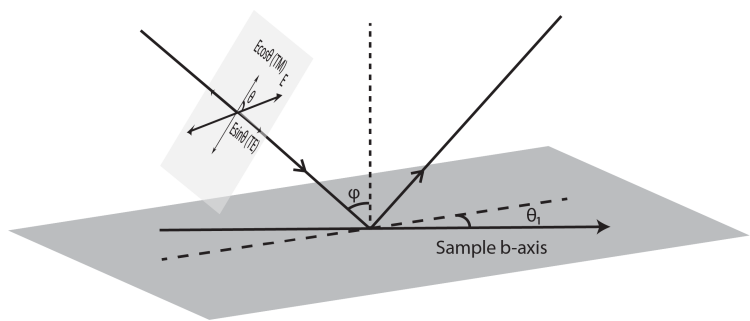

**Fig S1.** Ray diagram showing how rotation of polarization angle changes the in-plane TE and TM components of incident light for the reflectance setup.

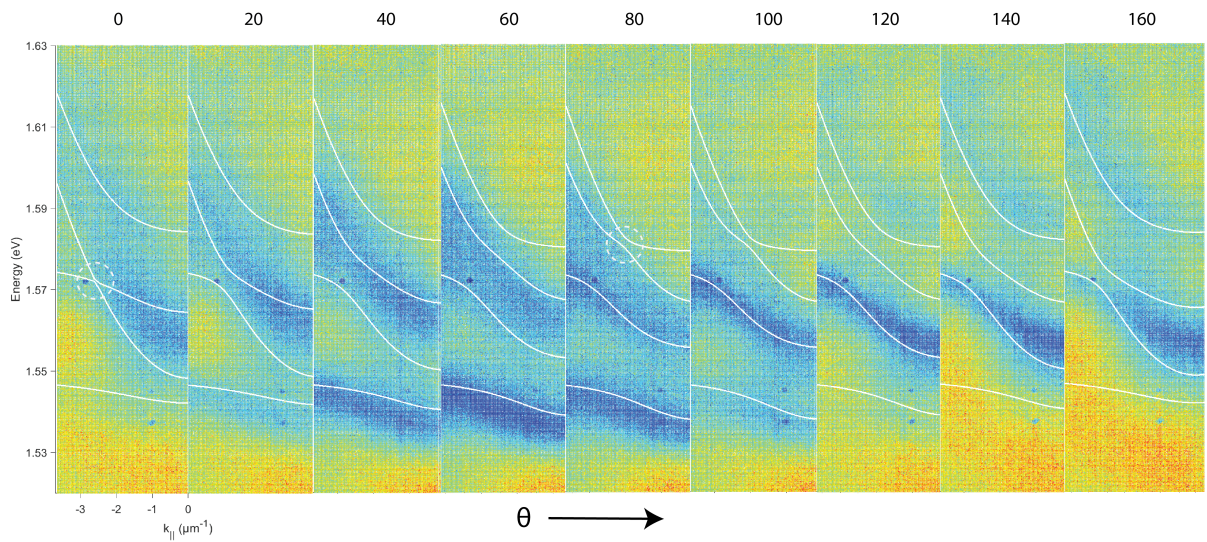

**Fig S2.** Polarization angle (θ) dependent, angle-resolved reflectance for 10nm ReS$_2$ in microcavity. White lines mark the polariton dispersions as obtained from a 4x4 coupled oscillator model.

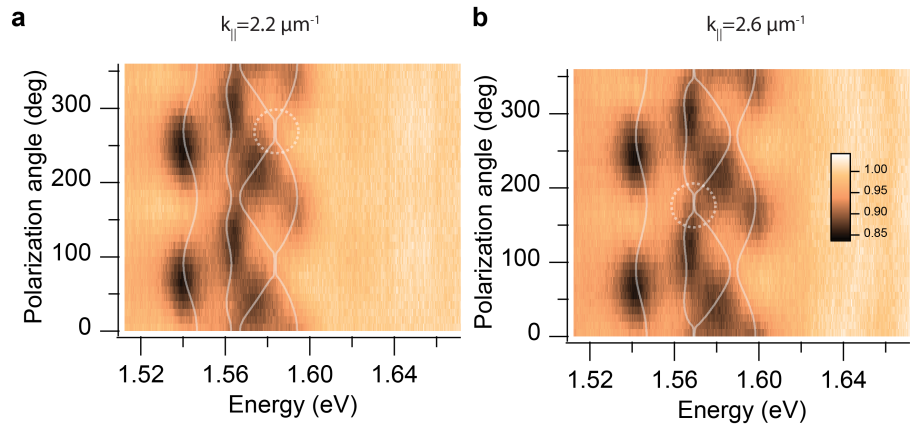

**Fig S3. a, b.** Experimental reflectance as a function of θ, for fixed k$_\parallel$ corresponding to where energy detuning with X$_2$ is zero for TM and TE cavity mode respectively. White translucent curves show the corresponding cross sections of the 3-D eigenenergy surfaces plotted in Fig 1b.

**Supplementary Note 1**

Cavity exciton-polaritons, which are hybrid light-matter quasiparticles formed by the coupling of excitons with photons inside high-quality optical resonators[43–45], offer an ideal platform to observe non-Hermitian physics due to the inherent losses in the system. The Hamiltonian of a cavity polariton is a 2x2 coupled oscillator matrix:

$$\begin{pmatrix} E_{cav} & g \\ g & E_{ex} \end{pmatrix}$$

Where typically $E_{cav}$ and $E_{ex}$, the cavity and exciton energies, are complex quantities carrying information about the cavity photon and exciton lifetimes. To probe EPs present in the eigenenergy manifold, the system should have the ability to realize different values of energy detuning ($\Delta = E_{cav} - E_{ex}$) and coupling strength ($g$). For $\Delta = 0$ and $g =$

$\sqrt{(\Gamma_{cav} - \Gamma_{ex})/2}$, where $\Gamma$ are the imaginary, dissipative parts of $E_{cav}$ and $E_{ex}$, the eigenenergies of the polariton modes will coalesce and give rise to an EP.